\documentclass[conference]{IEEEtran}
\IEEEoverridecommandlockouts

\usepackage{siunitx}
\usepackage{makecell}
\usepackage{ragged2e, microtype}
\usepackage{subcaption}
\usepackage{pifont}
\usepackage{amssymb}
\usepackage{float}
\usepackage{array}

\usepackage{url}
\usepackage{etoolbox}
\usepackage{soul} 
\usepackage{graphicx}
\usepackage{amsmath}
\usepackage{algorithm}
\usepackage{algpseudocode}
\usepackage{tabularx}
\usepackage[table]{xcolor}

\definecolor{lightgray}{rgb}{0.8, 0.8, 0.8}

\usepackage[left=0.63in,
            right=0.65in,
            top=0.75in,
            bottom=1in]{geometry}
            
\newtoggle{finalPaper}
\setstcolor{red}

\toggletrue{finalPaper} 

\iftoggle{finalPaper} {

	\newcommand{\rmvtxt}[1]{}
 	
	}{

	\newcommand{\rmvtxt}[1]{\st{#1}}
 	
}
\begin{document}

\newcommand{\yestick}{{\color{olive}\ding{51}}}
\newcommand{\notick}{{\color{red}\ding{55}}}

\title{Transfer Learning in Pre-Trained Large Language Models for Malware Detection Based on System Calls}

\author{{Pedro Miguel {S\'anchez S\'anchez}$^{*1}$, Alberto {Huertas Celdr\'an}$^{2}$, G\'er\^ome Bovet$^{3}$, Gregorio {Mart\'inez P\'erez}$^{1}$
}

\thanks{$^{*}$Corresponding author.}

\thanks{$^{1}$Pedro Miguel S\'anchez S\'anchez and Gregorio Mart\'inez P\'erez are with the Department of Information and Communications Engineering, University of Murcia, 30100 Murcia, Spain {\tt\small (pedromiguel.sanchez@um.es; gregorio@um.es)}.}%
\thanks{$^{2}$Alberto Huertas Celdr\'an are with the Communication Systems Group (CSG) at the Department of Informatics (IfI), University of Zurich UZH, 8050 Zürich, Switzerland {\tt\small (e-mail: huertas@ifi.uzh.ch).}}
\thanks{$^{3}$G\'{e}r\^{o}me Bovet is with the Cyber-Defence Campus within armasuisse Science \& Technology, 3602 Thun, Switzerland {\tt\small (gerome.bovet@armasuisse.ch)}.}
\thanks{This work has been partially supported by \textit{(a)} the Swiss Federal Office for Defense Procurement (armasuisse) with the DATRIS and CyberForce projects, and \textit{(b)} the University of Zürich UZH, and \textit{(c)} the strategic project CDL-TALENTUM from the Spanish National Institute of Cybersecurity (INCIBE) by the Recovery, Transformation, and Resilience Plan, Next Generation EU.}}

\maketitle

\begin{abstract}
In the current cybersecurity landscape, protecting military devices such as communication and battlefield management systems against sophisticated cyber attacks is crucial. Malware exploits vulnerabilities through stealth methods, often evading traditional detection mechanisms such as software signatures. The application of ML/DL in vulnerability detection has been extensively explored in the literature. However, current ML/DL vulnerability detection methods struggle with understanding the context and intent behind complex attacks. Integrating large language models (LLMs) with system call analysis offers a promising approach to enhance malware detection. This work presents a novel framework leveraging LLMs to classify malware based on system call data. The framework uses transfer learning to adapt pre-trained LLMs for malware detection. By retraining LLMs on a dataset of benign and malicious system calls, the models are refined to detect signs of malware activity. Experiments with a dataset of over 1TB of system calls demonstrate that models with larger context sizes, such as BigBird and Longformer, achieve superior accuracy and F1-Score of approximately 0.86. The results highlight the importance of context size in improving detection rates and underscore the trade-offs between computational complexity and performance. This approach shows significant potential for real-time detection in high-stakes environments, offering a robust solution to evolving cyber threats.

\end{abstract}

\begin{IEEEkeywords}
Large Language Model, Malware Detection, System Call, Transfer Learning, Cybersecurity
\end{IEEEkeywords}

\section{Introduction}
\label{sec:intro}

In the contemporary landscape of cybersecurity, particularly within the defense sector, safeguarding military assets represents a critical concern. Military devices, often connected to different networks, are frequent targets of sophisticated cyberattacks \cite{steingartner2021cyber}. Malware, a predominant threat vector, exploits vulnerabilities through deceptive and often undetected methods such as zero-day vulnerabilities \cite{theron2019autonomous}. The detection and neutralization of such threats are paramount, not only to ensure the operational integrity of military systems but also to protect critical security interests.

Incorporating artificial intelligence (AI) into cybersecurity represents a transformative shift in how defenses are conceptualized and deployed. AI technologies, through their ability to process and analyze vast amounts of data at high performance, provide a significant advantage in identifying and responding to cyber threats \cite{taddeo2019trusting}. Machine Learning and Deep Learning (ML/DL) algorithms, a subset of AI, are particularly adept at learning from and adapting to new information, thereby continuously improving threat detection models \cite{sarker2020cybersecurity}. This dynamic capability is crucial in an environment where threat actors continually evolve their methods to bypass conventional security measures.

Recent advancements in artificial intelligence (AI), specifically through the development of large language models (LLMs) for Natural Language Processing (NLP), have introduced promising new methodologies for enhancing cybersecurity measures \cite{motlagh2024large}. Some application scenarios involve spam filtering, smart contract auditing, or code analysis, among others. Among these solutions, transfer learning emerges as a particularly efficacious strategy. This approach leverages the extensive knowledge base of LLMs, originally trained on vast datasets, to adapt them to specialized domains with relatively sparse data, such as malware detection in military systems \cite{kucharavy2023fundamentals}.

Despite the advances in the application of LLMs in the cybersecurity field, some challenges remain open: (i) adapting LLMs to effectively process non-linguistic data such as system calls, which may not inherently fit the natural language processing paradigms these models were originally designed for; (ii) ensuring real-time detection capabilities within operational constraints, as LLMs can be computationally intensive and may not suit the time-sensitive needs of military operations; (iii) handling the trade-offs between context length and detection accuracy, as longer context may improve accuracy but also increases computational demands and latency; (iv) scaling the models to handle diverse and evolving attack vectors without extensive manual updates.

To solve the previous challenges, the main contributions of this work are:
\begin{itemize}
    \item A framework that utilizes system call traces from the device and an LLM model for the detection and classification of malware samples. The framework leverages pre-trained LLMs to refine their knowledge and add a classification layer on top of the model.
    
    \item The proposed framework is validated using a real dataset containing +1TB system calls collected in an RPi3-based spectrum sensor \cite{MalwSpecSys2022}. Several LLMs are tested, and their performance is compared according to factors such as the attention window. The best results are achieved with BigBird and Longformer models as both of them obtain $\approx 0.86$ in typical classification metrics such as accuracy or F1-Score
    
    \item A discussion analyzing the trade-off between detection speed and performance. It analyzes the processed data and the achieved results, remarking how LLM-based outputs can be later processed to achieve higher performance.
    
\end{itemize}

The remainder of this article is organized as follows. Section~\ref{sec:related} reviews the current state-of-the-art in language model-based syscall analysis and LLMs in cybersecurity. Section~\ref{sec:design} describes the design of the proposed solution employing LLMs for malware detection. Section~\ref{sec:validation} validates the solution in a real-world syscall dataset containing malware execution samples. Section~\ref{sec:discussion} reflects about the achieved results and the advantages and drawbacks of the solution. Finally, Section~\ref{sec:conclusions} depicts the conclusions obtained and the future work.

\section{Related Work}
\label{sec:related}

This section reviews the current state-of-the-art on the application of LLMs, and other NLP methods, in the area of anomaly detection, security, and malware detection using system calls as inputs.

N-gram-based methods for system call processing have been utilized as a baseline in the field of cybersecurity and malware detection \cite{hubballi2011sequencegram}. More recently, n-gram techniques have been applied as feature extraction approaches for ML/DL-based methods \cite{ali2020malgra, celdran2022privacy}. However, their primary limitation lies in the lack of contextual understanding, as n-grams consider only fixed-length sequences and cannot capture longer dependencies or complex behaviors that span beyond the defined n-size window. This often results in a higher rate of false positives and negatives, particularly with sophisticated malware that employs evasion techniques or mimics benign behaviors.

Regarding LLMs for syscall processing, the authors of \cite{fournier2023language} compared LSTM, Transformers, and Longformers (a Transformer variant with lower complexity) with 4-gram statistical approaches for syscall classification. They published a dataset covering the system calls generated by seven behaviors monitored in two million web requests. The neural network-based solutions achieved better performance, especially for LSTM and Transformer. They remarked on the challenges of real-time execution of complex language models.

Focused on intrusion detection, the authors of \cite{kim2016lstm} proposed an ensemble method based on LSTM models that achieved a 0.928 AUC (Area Under Curve), 0.16 FAR (False Alarm Rate) for 0.90 (Detection Rate), in relevant literature syscall datasets such as AFDA-LD and KDD98 \cite{khraisat2019survey}. In \cite{almodovar2022can}, authors demonstrated that a Longformer-based approach improves previous BERT-based solution \cite{chen2022bert} in system log anomaly detection due to its larger context size. Data leveraged in this study contains logs from HFDS and Thunderbird applications and the BGL supercomputer system.


As can be seen in the related work comparison in \tablename~\ref{tab:related} and recent literature reviews confirm \cite{motlagh2024large}, LLMs have not been extensively applied in syscall data in the realm of intrusion or malware detection, remaining a field requiring considerable research efforts.

\begin{table}[h!]
\centering
\caption{LLMs and NLP Methods in Syscall Processing and/or Cybersecurity}
\label{tab:related}
\begin{tabular}{p{0.55cm}p{0.55cm}p{2cm}p{1.5cm}p{2cm}}
\hline
\textbf{Work} & \textbf{Year} & \textbf{Approach} & \textbf{Model} & \textbf{Results} \\ \hline
\cite{fournier2023language} & 2023 & Syscall behavior classification & LSTM, Transformer, Longformer & Improved performance over 4-gram \\ \hline
\cite{kim2016lstm} & 2016 & Develop intrusion detection system & LSTM ensemble & AUC 0.928, FAR 0.16 \\ \hline
\cite{almodovar2022can} & 2022 & Application log anomaly detection & Longformer-based & Larger context size improved BERT and LSTM models \\ \hline
This & 2024 & Malware classification & Longformer, BigBird & +0.86 accuracy, precision, recall, F1-score. Context size analysis \\ \hline
\end{tabular}
\end{table}

\section{Design}
\label{sec:design}

\begin{figure}[htpb!]
    \centering
    \includegraphics[width=0.9\columnwidth]{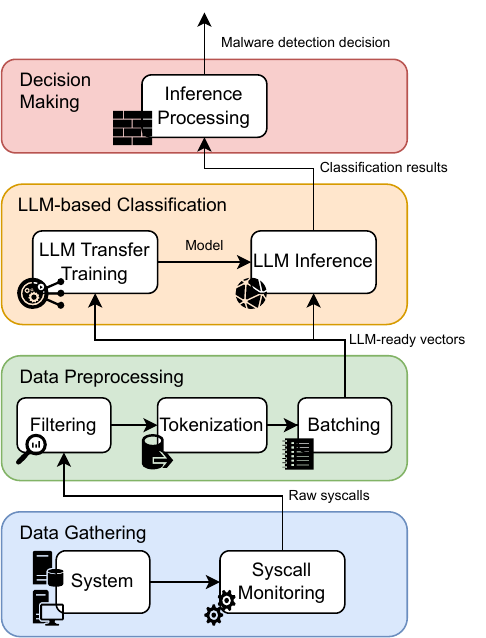}
    \caption{LLM-based malware detection framework}
    \label{fig:framework}
\end{figure}

This section details the design of the framework for malware detection and classification based on system calls. \figurename~\ref{fig:framework} describes the proposed framework and the interaction between its components.

\subsection{Data Gathering}

This module is responsible for capturing system calls generated by applications running on the device. System calls (syscalls), which represent the interface between user applications and the operating system, provide a rich source of information about application behavior. This module employs a syscall monitor at the operating system level, which hooks into the syscall dispatcher to log relevant information such as syscall identifiers, arguments, return values, and timestamps. The collected data is crucial for understanding the actions performed by applications, enabling the identification of potentially malicious behavior patterns. By continuously monitoring and recording syscalls, this module ensures that comprehensive and detailed data is available for subsequent analysis. Once syscalls are captured and logged, they can either be sent to an external server for processing or processed on-device, depending on the device processing capabilities.

\subsection{Data Preprocessing}
The preprocessing module prepares the captured syscall data for analysis by the LLM. It primarily focuses on data tokenization and batching. Data tokenization involves converting raw syscall data into a format that the LLM can process, typically by mapping the text of each syscall and its attributes to unique tokens. These tokens preserve the sequence and contextual information necessary for accurate analysis. Note that the tokenizer should be configured according to the LLM deployed later in the pipeline. Once tokenized, the data is organized into batches to facilitate efficient processing by the LLM. Batching helps manage memory usage and speeds up the computation by allowing parallel processing of multiple data sequences. This streamlined preprocessing ensures that the data is in an optimal state for subsequent classification by the LLM.

\subsection{LLM-based Classification}

At the core of the framework is the LLM-based classification module. This module performs the critical task of classifying syscalls to determine whether they indicate malware activity. The LLM analyzes the sequences and contexts of incoming syscall data and classifies them as either normal or malicious based on the learned patterns. Pre-trained LLMs are leveraged in this module, adapting their final layers for classification and applying a transfer learning process using the syscall data.

\subsubsection{Model Training}
This submodule is responsible for adapting the LLM to the specific task of syscall classification. The training involves supervised learning using labeled datasets comprising examples of both benign and malicious syscall patterns. During training, the model learns to differentiate between these patterns by adjusting its parameters. The goal is to minimize classification errors and improve the model accuracy in distinguishing between benign and malicious behaviors.

\subsubsection{Model Inference}
Once trained, the inference submodule uses the model to perform real-time or batch analysis of incoming syscall data. Based on the learned patterns, it classifies the data as either malicious or benign. This real-time analysis ensures timely detection and response to potential threats.

\subsection{Decision Making}

The Decision Making module processes outputs from the LLM to determine the presence of malware, utilizing probabilities to evaluate the likelihood of malicious activity. User-defined thresholds adjust sensitivity and specificity, allowing for tailored detection settings. Lower thresholds increase sensitivity but may result in more false positives, while higher thresholds enhance specificity, potentially missing some threats. Beyond threshold-based decisions, the module can aggregate LLM outputs over time or across devices to detect broader patterns. Additionally, the probabilities can train a secondary model, such as a logistic regression or decision tree, to refine classifications. This approach enhances precision by leveraging the LLM scores. The module integrates these techniques to ensure accurate and appropriate responses, such as alerting users, quarantining processes, or blocking malicious activities, thereby enhancing the framework overall reliability and effectiveness.

\section{Validation}
\label{sec:validation}

This section describes the experiments performed to validate the proposed framework. This validation focused on the verification of the capabilities of LLMs to detect and classify malware samples, as the data gathering part of the framework has been implemented in the literature previously \cite{celdran2022privacy}.

\subsection{MalwSpecSys Dataset}

The selected dataset for validation is MalwSpecSys \cite{MalwSpecSys2022}. This dataset models the internal behavior of an IoT spectrum sensor belonging to the ElectroSense platform \cite{rajendran2017electrosense}, consisting of a Raspberry Pi 3 with a software-defined radio kit. The behavior was monitored when it was functioning normally and under different malware attacks. Syscalls were collected using the \textit{perf trace} tool.

This dataset contains +1 TB of information collected during several days. The data consists of raw syscall information with the format [\textit{timestamp}, \textit{process}, \textit{PID}, \textit{syscall}], and is divided into one file every 10 seconds. 

Regarding the malware samples deployed during data collection, they have different varieties of common cyberattacks:

\begin{itemize}
    \item \textit{Bashlite} \cite{bashlite}, a well-known botnet family targeting IoT devices. It is capable of launching distributed denial-of-service (DDoS) attacks, executing arbitrary shell commands, and enlisting infected devices into a botnet.

    \item \textit{TheTick} \cite{thetick}, a backdoor used to control bots from a remote location through a server utilizing a remote shell and retrieving data from targeted devices

    \item \textit{Bdvl} \cite{bdvl}, a rootkit with very wide functionality. It ranges from hidden backdoors that allow multiple connection methods to keylogging and stealing passwords and files.

    \item \textit{RansomwarePoC} \cite{ransomware}, an example of ransomware implementation in Python with full encryption capabilities. Only the C\&C functionality is missing.
\end{itemize}

As the syscalls generated by normal behavior and each attack are known, the problem is addressed as a classification task with five different labels, one for normal behavior and four for the different malware.

For preprocessing, the \textit{timestamp}, \textit{process}, and \textit{PID} are removed from the dataset, maintaining only the syscalls ordered in time and without call parameters. Besides, numerous \textit{nanosleep} syscalls are removed as they do not provide useful information about the system activities. After processing, each file contains $\approx$23k syscalls. These sequences are the ones that are fed as input in the LLM models. The number of syscalls per sequence varies according to the maximum context length of the models.

\begin{table*}[ht!]
\centering
\caption{Classification Results of LLMs}
\label{tab:llm_comparison}
\begin{tabular}{p{1.4cm}cp{1cm}p{1cm}p{0.8cm}p{1.2cm}p{1cm}p{0.8cm}}
\hline
\textbf{Model} & \textbf{Context Size} & \textbf{Accuracy} & \textbf{Precision} & \textbf{Recall} & \textbf{F1-Score} & \textbf{Kappa} & \textbf{MCC}\\ \hline
BERT & 512 & 0.6772 & 0.8200 & 0.6596 & 0.6465 & 0.5504 & 0.6024 \\ \hline
DistilBERT & 512 & 0.6289 & 0.7100 & 0.6181 & 0.5930 & 0.4786 & 0.5379 \\ \hline
GPT-2 & 1024 & 0.6944 & 0.7986 & 0.6865 & 0.6808 & 0.5792 & 0.6123 \\ \hline
\rowcolor{lightgray!65} BigBird & 4096 & 0.8667 & 0.8754 & 0.8668 & 0.8688 & 0.8298 & 0.8311 \\ \hline
Longformer & 4096 & 0.8616 & 0.8696 & 0.8614 & 0.8621 & 0.8232 & 0.8250 \\ \hline
Mistral & 8192 & 0.5817 & 0.6112 & 0.6462 & 0.6242 & 0.4754 & 0.4798 \\ \hline
\end{tabular}
\end{table*}

\subsection{LLMs for Transfer-Learning}

Once the data is ready, diverse pre-trained LLMs are tested. A final layer with five neurons is added to each model to adapt it to the classification task, having one neuron per class. The base tokenizer of each LLM is employed to preprocess the text data and make it appropriate for the model format.

Then, five training epochs are executed to personalize the LLM to the new task. For implementation purposes, Huggingface's Trasnsformers library is employed \cite{wolf2019huggingface}. As the optimizer function, AdamW with a $1e-5$ learning rate is employed in all cases. The experiments are executed in a compute node leveraging an AMD EPYC 7742 CPU and an NVIDIA A100 40GB GPU. Although more GPUs were available, only one was used in order to be representative of real-world complete framework deployments involving LLM retraining where exhaustive computing resources are not available.

The selected LLMs for validation are open-source and commonly utilized in transfer learning problems. Besides, the chosen models remain relatively small, avoiding using models with billions of parameters. Each model is characterized by its context size, impacting its ability to handle long sequences of text effectively.
\begin{itemize}
    \item \textit{BERT} (Bidirectional Encoder Representations from Transformers) has a context size of 512 tokens. It is widely used for various natural language understanding tasks due to its bidirectional attention mechanism, which allows it to capture context from both directions.
    \item \textit{DistilBERT}, a smaller and faster version of BERT, also with a context size of 512 tokens. It maintains 97\% of BERT's language understanding capabilities while being more efficient, making it suitable for resource-constrained environments.
    \item \textit{GPT-2} (Generative Pre-trained Transformer 2) supports a context size of 1024 tokens. It is known for its strong text generation capabilities, leveraging its autoregressive nature to produce coherent and contextually relevant text.
    \item \textit{BigBird}, with a context size of 4096 tokens, extends the Transformer architecture to handle longer sequences efficiently. It combines sparse attention mechanisms to manage large contexts, making it ideal for tasks requiring extensive contextual information.
    \item \textit{Longformer} supports context sizes of up to 16384 tokens depending on the implementation. It employs a combination of local and global attention mechanisms to efficiently process long documents, particularly useful for tasks like document classification and summarization. The model tested uses a context size of 4096.
    \item \textit{Mistral Small}, designed with an impressive maximum context size of 128,000 tokens, utilizes advanced techniques like sliding window attention, making it ideal for tasks involving extensive context and long-form content generation. The model tested uses a context size of 8192.
\end{itemize}

Several classification metrics based on the confusion matrix are calculated to measure classification performance. The confusion matrix is a table used to describe the performance of a classification model on a set of data for which the true values are known. It classifies predictions into four categories: True Positives (TP), True Negatives (TN), False Positives (FP), and False Negatives (FN).

\begin{itemize}
    \item Accuracy evaluates the overall correctness of the model and is calculated as the ratio of correctly predicted observations to the total observations.
\small
\begin{equation}
    \text{Accuracy} = \frac{\text{TP} + \text{TN}}{\text{TP} + \text{TN} + \text{FP} + \text{FN}}
\end{equation}
\normalsize

\item Precision or True Positive Rate (TPR) assesses the accuracy of positive predictions made by the model.
\small
\begin{equation}
    \text{Precision or TPR}= \frac{\text{TP}}{\text{TP} + \text{FP}}
\end{equation}
\normalsize

\item Recall (or sensitivity) measures the model ability to detect positive samples.
\small
\begin{equation}
    \text{Recall} = \frac{\text{TP}}{\text{TP} + \text{FN}}
\end{equation}
\normalsize

\item The F1-Score is the harmonic mean of precision and recall, providing a balance between the two by considering both false positives and false negatives.
\small
\begin{equation}
    \text{F1-Score} = 2 \times \frac{\text{Precision} \times \text{Recall}}{\text{Precision} + \text{Recall}}
\end{equation}
\normalsize

\item Cohen’s Kappa statistic adjusts accuracy by accounting for the possibility of agreement occurring by chance.
\footnotesize
\begin{equation}
    \kappa = \frac{2 \times (TP \times TN - FN \times FP)}{(TP+FP)\times(FP+TN)+(TP+FN)\times(FN+TN)}
\end{equation}
\normalsize

\item Matthews Correlation Coefficient (MCC) is a reliable statistical rate that produces a high score only if the prediction obtained good results in all of the four confusion matrix categories (TP, TN, FP, FN), proportionally.
\footnotesize
\begin{equation}
    \text{MCC} = \frac{\text{TP} \times \text{TN} - \text{FP} \times \text{FN}}{\sqrt{(\text{TP} + \text{FP})(\text{TP} + \text{FN})(\text{TN} + \text{FP})(\text{TN} + \text{FN})}}
\end{equation}
\normalsize

\end{itemize}

\tablename~\ref{tab:llm_comparison} details the results achieved for each one of the tested LLMs. It can be seen how the increase in the context size improves the model performance sequentially until it reaches 4096 tokens with Longformer and BigBird models. After that limit, the Mistral model with 8192 context size performed worse than other models because Normal and TheTick behaviors were misclassified. In contrast, Bashlite, Bdvl, and RansomwarePoC behaviors were reliably classified with +0.96 TPR.

\begin{figure}[htpb!]
    \centering
    \includegraphics[width=\columnwidth]{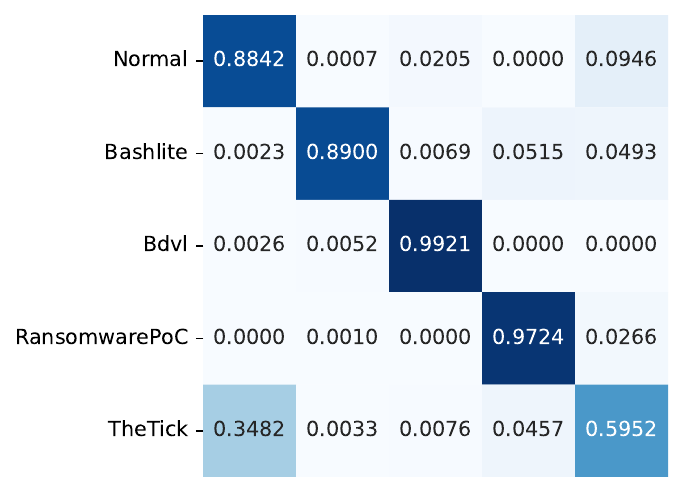}
    \caption{BigBird Confusion Matrix}
    \label{fig:BigBird}
\end{figure}

\figurename~\ref{fig:BigBird} shows the confusion matrix for the BigBird model, the best-performing model. It illustrates the classification performance across the different malware families and normal behavior. The model achieved a high TPR for normal behavior at 0.8842 but misclassified 0.205 as Bdvl and 0.946 as TheTick. For Bashlite, the accuracy was 0.8900, with minimal misclassifications. Bdvl was correctly identified with a TPR of 0.9921. RansomwarePoC showed a TPR of 0.9724 with minor misclassifications. TheTick had a lower TPR of 0.5952, with 0.3482 misclassified as normal behavior. The results highlight the effectiveness of BigBird in identifying specific threats and the need for further refinement in detecting TheTick. 

\section{Discussion}
\label{sec:discussion}

The results of the validation underscore the importance of context size in enhancing the performance of LLMs for malware classification tasks. It is observed that models with larger context sizes, such as BigBird and Longformer, achieved higher accuracy and better classification metrics compared to those with smaller context sizes, like BERT and DistilBERT. Specifically, BigBird, with a context size of 4096 tokens, demonstrated superior performance in accurately classifying the syscall patterns of various malware types, as evidenced by its high TPR across multiple classes. The main reason for the performance improvement with the context size increase is the large number of syscalls being generated per second. The dataset contains roughly 23,000 syscalls every 10 seconds, which is a rate of +2,000 syscalls per second. Therefore, models with context sizes of 512 or 1024 are not able to process the syscalls generated every second. During that short period, the malware samples might not perform any operations, even if they are active in the device. 

However, the results also reveal some trade-offs between context size and performance. While increased context size generally improves the model ability to capture and utilize extended sequences of syscalls, it also introduces challenges. For instance, the Mistral model, despite its capacity to handle up to 128,000 tokens, showed reduced performance at a context size of 8192 tokens. This drop in performance can be attributed to the increased complexity and potential overfitting when dealing with excessively large contexts without sufficient data diversity or volume to support such detailed analysis.

The misclassification of TheTick as normal behavior highlights a critical area for further refinement. This suggests that while the model can effectively utilize larger contexts to improve detection rates for certain malware types, it may still struggle with specific patterns or classes that require more nuanced differentiation. This finding points to the need for balanced context sizes that maximize information utility without overwhelming the model discriminative capabilities.

Another significant aspect to consider is how LLM predictions can be aggregated to enhance detection accuracy. Advanced detection systems can benefit from ensemble methods, where predictions from multiple models with varying context sizes are combined to form a consensus decision. This approach leverages the strengths of each model, potentially offsetting their individual weaknesses. For instance, predictions from BERT and DistilBERT, which are efficient and effective for shorter contexts, can be combined with those from BigBird and Longformer for a more comprehensive analysis. Weighted averaging, voting mechanisms, or more sophisticated ensemble techniques like stacking can be employed to aggregate predictions, thereby improving overall detection robustness.

Works in the literature dealing with similar scenarios, such as \cite{celdran2023intelligent}, have achieved higher detection rates in the malware evaluated in this work using kernel events as data source. Concretely, using ten-second windows, a 0.94 average F1-score was achieved. However, note that the 4096 context size in the best-performing LLMs of this work (Longformer and BigBird) allows the processing of around one second of syscall data per evaluation. Therefore, the aggregation of the predictions would be necessary to achieve higher performance with similar time windows.

Integrating LLM predictions with additional context-specific features, such as temporal patterns of syscalls or correlation with network activities, could further enhance the detection framework. This multi-faceted approach allows for the consideration of not just static syscall patterns but also their dynamic behavior over time, providing a richer context for identifying sophisticated malware that employs evasion techniques.

\section{Conclusions and Future Work}
\label{sec:conclusions}

This work proposes a malware detection framework based on LLM transfer learning. It leverages LLMs ability to process and analyze sequences of syscalls. The data preprocessing module ensures the syscalls are tokenized and batched for efficient processing. The core of the framework is the LLM-based classification module, which analyzes the syscall sequences and classifies them as benign or malicious. The decision-making module processes the classification results, applying thresholds to determine the presence of malware.

During validation, pre-trained LLMs, including BERT, DistilBERT, GPT-2, BigBird, Longformer, and Mistral, were adapted with an additional classification layer for syscall analysis. Using a dataset of over 1TB of system calls from a Raspberry Pi 3-based spectrum sensor, that captured both normal and malicious activities for training and validation.

The results showed that models with larger context sizes, such as BigBird and Longformer, achieved better classification metrics compared to those with smaller context sizes. Specifically, BigBird and Longformer, with a context size of 4096 tokens, demonstrated superior performance, achieving accuracy and F1-Score of approximately 0.86. However, the Mistral model, despite its ability to handle up to 128,000 tokens, performed worse at a context size of 8192 tokens. This reduced performance is attributed to the complexity and potential overfitting associated with excessively large contexts.

Future work considers applying model quantization techniques to deploy the LLMs in the devices themselves, avoiding the need for external processing. Another future direction is adapting the LLMs for anomaly detection using only benign data, as this could potentially allow the detection of zero-day attacks not seen previously. Additional context-specific features, such as temporal patterns of syscalls and correlation with network activities, will be integrated to provide a more comprehensive analysis of potential threats. Real-time detection capabilities will be enhanced to meet operational constraints, particularly in military applications.

\bibliographystyle{unsrt}
\bibliography{references}

\end{document}